\documentclass[twocolumn,preprintnumbers,amsmath,amssymb,superscriptaddress]{revtex4-2}

\usepackage{amsmath}
\usepackage{float}
\usepackage{amsfonts}
\usepackage{amssymb}
\usepackage{subfiles}
\usepackage[colorlinks,linkcolor=red,citecolor=blue,breaklinks=true]{hyperref}
\usepackage{graphicx}% Include figure files
\usepackage{dcolumn}% Align table columns on the decimal point
\usepackage{bm}% bold math
\usepackage{amssymb,amsmath,color}
\usepackage{booktabs}
\usepackage{amsthm}
\usepackage{dsfont}
\usepackage{tcolorbox}
\usepackage{cases}% http://ctan.org/pkg/cases
\usepackage{physics}
\usepackage{subfiles}

\usepackage{braket}
\usepackage{bbold}
\usepackage{xcolor}
\usepackage{ulem} % for crossing out
\usepackage{txfonts,times}
\usepackage{xcolor}

\begin{document}
\date{\today}

\title{Certifying semi-device-independent security via wave-particle duality experiments}

\author{Chithra Raj}
\email{These authors have contributed equally to this work.}
\affiliation{International Centre for Theory of Quantum Technologies,
University of Gdańsk, Jana Bazynskiego 8, 80-309 Gdańsk, Poland}

\author{Tushita Prasad}
 \email{These authors have contributed equally to this work.}
\affiliation{%
International Centre for Theory of Quantum Technologies,
University of Gdańsk, Jana Bazynskiego 8, 80-309 Gdańsk, Poland}%

\author{Anubhav Chaturvedi}
%\email{anubhav.chaturvedi@pg.edu.pl}
\affiliation{International Centre for Theory of Quantum Technologies,
University of Gdańsk, Jana Bazynskiego 8, 80-309 Gdańsk, Poland}
\affiliation{Faculty of Applied Physics and Mathematics, Gda{\'n}sk University of Technology, Gabriela Narutowicza 11/12, 80-233 Gda{\'n}sk, Poland}

\author{Lucas Pollyceno}
% \email{}
\affiliation{%
International Centre for Theory of Quantum Technologies,
University of Gdańsk, Jana Bazynskiego 8, 80-309 Gdańsk, Poland}%

\author{Daniel Spegel-Lexne}
%\email{}
\affiliation{Institutionen f\"{o}r Systemteknik, Link\"opings Universitet, 581 83 Link\"oping, Sweden}

\author{Santiago G\'omez}
%\email{}
\affiliation{Dipartimento di Fisica, Sapienza Universit\`a di Roma, Piazzale Aldo Moro 5, I-00185 Roma, Italy.}

\author{Joakim Argillander}
\affiliation{Institutionen f\"{o}r Systemteknik, Link\"opings Universitet, 581 83 Link\"oping, Sweden}

\author{Alvaro Alarc\'on}
\affiliation{Departamento de
Ingeniería Eléctrica y Electrónica, Facultad de Ingeniería, Universidad del Bío-Bío, Avenida Collao 1202, 4051381, Concepción Chile.}
%% \affiliation{Institutionen f\"{o}r Systemteknik, Link\"opings Universitet, 581 83 Link\"oping, Sweden}

\author{Guilherme B. Xavier}
\email{guilherme.b.xavier@liu.se}
	\affiliation{Institutionen f\"{o}r Systemteknik, Link\"opings Universitet, 581 83 Link\"oping, Sweden}

    \author{Marcin Pawłowski}
\affiliation{International Centre for Theory of Quantum Technologies,
University of Gdańsk, Jana Bazynskiego 8, 80-309 Gdańsk, Poland}

\author{Pedro R. Dieguez}
\email{pedro.dieguez@ug.edu.pl}
\affiliation{International Centre for Theory of Quantum Technologies,
University of Gdańsk, Jana Bazynskiego 8, 80-309 Gdańsk, Poland}

%##########################################################################
%##########################################################################

\begin{abstract}
Wave-particle duality is known to be equivalent to an entropic uncertainty relation based on the min- and max-entropies, which have a clear operational meaning in quantum cryptography. Here, we derive a connection between wave-particle relations and the semi-device-independent (SDI) security framework. In particular, we express an SDI witness entirely in terms of two complementary interferometric quantities: visibility and input distinguishability.
Applying a symmetry condition to the interferometric quantities, we identify a scenario in which the classical bound is violated and the security condition is met in wave-particle experiments with a tunable beam splitter. This enables the certification of non-classicality and the positivity of the key rate directly from complementary interferometric quantities.
Moreover, we perform a proof-of-principle experiment using orbital-angular-momentum encoded quantum states of light in a tunable interferometer, validating our theoretical predictions. Finally, we analyze an improved bound on the SDI security condition, effectively enlarging the parameter region where secure communication can be certified.
\end{abstract}

%##########################################################################
%##########################################################################

\maketitle

\section*{Introduction}

Bohr’s complementarity principle, a cornerstone of quantum theory, provides a unified framework in which matter and radiation can exhibit either wave-like or particle-like behavior, depending on the experimental context~\cite{dieguez2022experimental,saunders2005complementarity}.  
Its validity has been thoroughly tested in a wide range of systems—from double-slit experiments with light~\cite{TYoung,king2010matterless,tirole2023double} and matter waves~\cite{matterWave,zimmermann2008localization,liu2015einstein} to modern delayed-choice scenarios~\cite{Wheeler,Terno,kaiser2012entanglement,auccaise2012experimental,dieguez2022experimental}—demonstrating that interference arises when path information is unavailable, while the reveal of which-path information suppresses interference and yields particle-like detections. These results, consistent across diverse platforms, highlight the contextual nature of quantum phenomena and challenge classical assumptions of simultaneously well-defined “elements of reality” for incompatible observables~\cite{dieguez2018information,Lustosa20,dieguez2022experimental,lustosa2025emergence}.

In a Mach–Zehnder interferometer (MZI), wave-particle duality can be expressed through conceptually distinct relations depending on the experimental configuration~\cite{Jaeger1995,englert1996fringe}. When the first beam splitter is biased, the setup defines a predictive scenario where the trade-off is between path predictability and fringe visibility. In contrast, a balanced first beam splitter defines a retrodictive scenario, where input distinguishability replaces predictability as the particle-like quantity. More generally, if an environment interacts with the system inside the interferometer, wave-particle duality is characterized by path distinguishability obtained from environmental measurements~\cite{coles2014equivalence,coles2016entropic,coles2017entropic}. In all cases, the fundamental principle holds: increased knowledge of particle-like behavior reduces the visibility of interference fringes, and vice versa.

A substantial refinement of these ideas arises from the discovery that wave–particle duality relations are equivalent to optimal entropic uncertainty relations formulated with min- and max-entropies~\cite{coles2014equivalence,coles2016entropic}.  
This equivalence lends an operational, information-theoretic meaning to the trade-off between the interferometric quantities: the same entropic quantities that limit simultaneous wave and particle manifestations also bound an eavesdropper’s optimal guessing probability and, therefore, the secrecy achievable in quantum key distribution (QKD)~\cite{scarani2009security,berta2010uncertainty,mizutani2017information,zhang2023quantum}. This equivalence was recently confirmed in a photonic experiment using orbital angular momentum (OAM) quantum states of light, where the equivalence between wave-particle duality and entropic uncertainty relations was demonstrated on a versatile, reconfigurable platform based on few-mode optical fibers and photonic lanterns~\cite{spegel2024experimental}. As a result, wave-particle duality emerges not merely as a conceptual pillar but as a practical resource for quantum information science.

Complementarity plays a key role in both device-dependent and device-independent quantum cryptography. In the device-dependent regime, it underlies the security of protocols like BB84, where mutually unbiased bases ensure that eavesdropping induces detectable disturbances~\cite{koashi2009simple}. Delayed-choice-inspired cryptographic protocol~\cite{Ardehali1996} similarly uses flexible measurement settings to enforce security through wave-particle duality.
In the device-independent (DI) paradigm, security relies solely on Bell inequality violations~\cite{Belltosecurity,Belltosecurity2,DI}, offering strong guarantees but requiring stringent experimental conditions~\cite{DI}. To mitigate these challenges, complementarity-based DI protocols have been proposed for more practical scenarios~\cite{zhang2023quantum}, and broader operational perspectives on complementarity have also been explored~\cite{saha}.
Given the experimental limitations of DI protocols, the semi-device-independent (SDI) framework—requiring only a bound on system dimension~\cite{pawlowski2011semi}—offers a promising alternative. While complementarity has been linked to both device-dependent and fully device-independent cryptographic frameworks, its role in the semi-device-independent (SDI) scenario remains largely unexplored.

In this work, we aim to bridge this gap by establishing a connection between wave-particle duality relations and semi-device-independent (SDI) security witnesses.  We derive a method to express an SDI witness in terms of interferometric quantities—visibility and input distinguishability—capturing the operational features of complementarity across arbitrary experimental contexts. 
We present a particular mapping to assess a region in the visibility-distinguishability plane where both the classical bound and the security threshold can be violated via wave-particle duality experiments with a final tunable beam splitter (TBS), thereby demonstrating non-classicality and key-rate positivity with purely interferometric data.   
Moreover, we analyze a better bound on the SDI security condition, which enlarges the accessible security region.
Finally, using a fiber-optical interferometer designed to be dynamically reconfigurable to measure the visibility and path distinguishability of OAM quantum states of light~\cite{spegel2024experimental}, we perform a proof-of-principle experiment that validates our main result.

%{\it Results:}
\section*{Results}
\label{sec: re}

\subsection{Theoretical framework}

We begin by outlining the theoretical framework underlying our analysis.
In general \textit{prepare-and-measure} (PM) scenarios, Alice prepares a message using a $d$-dimensional physical system from a set $a \in \{0, \ldots, N - 1\}$ and sends it to Bob through a potentially insecure channel. Bob performs a measurement chosen by $y \in \{0, \ldots, m - 1\}$ and obtains an outcome $b \in \{0, \ldots, k - 1\}$. The PM scenario is defined by the dimension $d$ and the tuple $(N, m, k)$. After many rounds, the observed statistics are given by the conditional probability distribution $\{P(b \mid a, y)\}_{b,a,y}$.
Such correlations are constrained by the available resources, especially the communication dimension $d$. When limited to classical resources of dimension $d$, the statistics must satisfy inequalities of the form
\begin{equation}
\sum_{a,y,b} w_{a,y,b} P(b \mid a, y) \leq C_d,
\end{equation}
where $C_d$ is the classical bound. These expressions are referred to as \textit{dimension witnesses}. Quantum systems can violate these bounds while still using dimension $d$, as the probabilities are governed by the Born rule
\begin{equation}
P(b \mid a, y) = \operatorname{Tr}(\rho_a M^b_y),
\end{equation}
where $\rho_a$ are $d$-dimensional quantum states and $M^b_y$ denotes Bob's measurement POVMs.

\begin{figure}
\centering
\includegraphics[width=1\columnwidth]{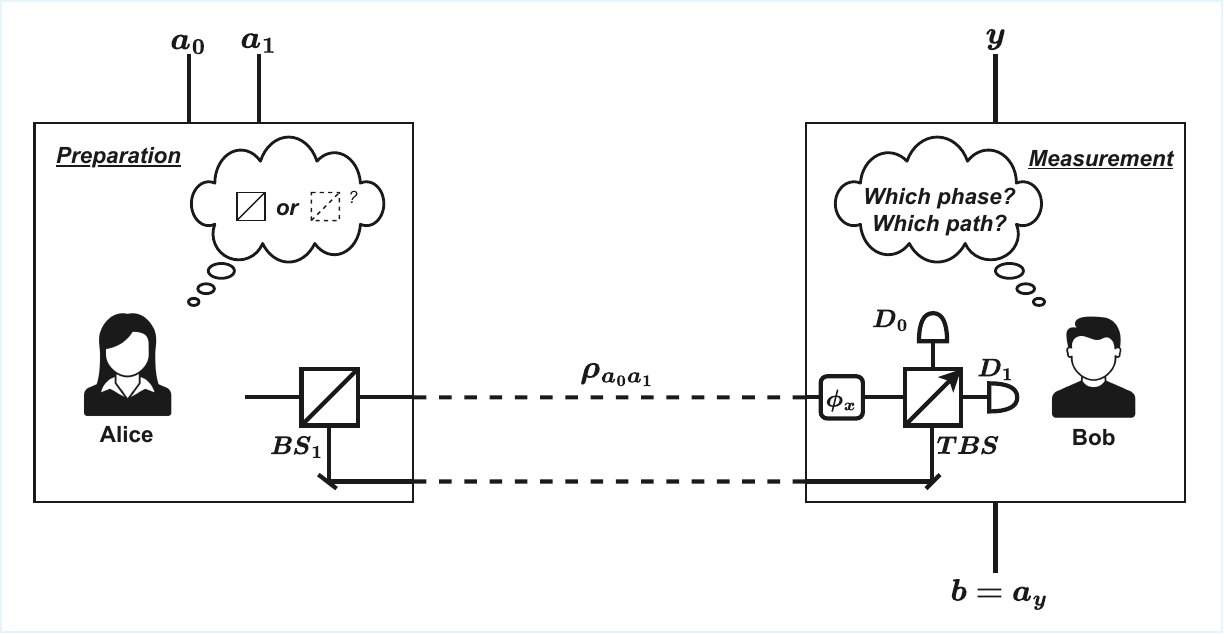}
\caption{\textbf{SDI Witness and Wave-Particle Games}: In the \textit{preparation stage}, Alice receives two classical bits \((a_0, a_1)\) and encodes them into a two-dimensional quantum state \(\rho_{a_0a_1}\). This encoding involves a choice of whether to insert an unbiased beam splitter (\(BS_1\)) on each computational basis preparation or not. In the \textit{measurement stage}, Bob receives the quantum state, after a phase shift $\phi_x$ is applied, and performs a measurement based on his chosen input. His setup includes a tunable beam splitter (TBS), which allows him to interpolate between particle-like and wave-like observables. Due to wave-particle duality and complementarity, Bob faces fundamental limits: he cannot simultaneously extract both which-path and which-phase information with arbitrary precision. These constraints, expressed in terms of interferometric visibility and path distinguishability, provide a basis for evaluating the SDI witness.}
\label{FigSDI}
\end{figure}

The SDI approach assumes a known dimension $d$, and a violation of the dimension witness certifies the system’s quantum nature without full device characterization. As demonstrated in Ref.~\cite{pawlowski2011semi}, such violations also ensure security in one-way QKD, as quantum messages cannot be copied like classical ones.
A particularly relevant scenario is $(4,2,2)$, as depicted in Fig.~\ref{FigSDI}, in which Alice selects one of four preparations, labeled by two bits $a_0a_1$, and Bob performs one of two binary measurements. The conditional probabilities are grouped into correlators $E_{a_0a_1,y} := P(b = 0 \mid a_0a_1, y)$, and the SDI witness takes the form:
\begin{equation}\label{eq:sdi_witness}
S = E_{00,0} + E_{00,1} + E_{01,0} - E_{01,1} - E_{10,0} + E_{10,1} - E_{11,0} - E_{11,1}.
\end{equation}
Classically, $S \leq 2$, but quantum strategies allow up to $S = 2\sqrt{2}$.
This witness is related to a quantum random access code (QRAC) task~\cite{pawlowski2011semi}, where Bob attempts to guess one of Alice’s bits $a_y$:
\begin{equation}\label{eq:suc_rac}
P_B = \frac{1}{8}\sum_{a_0, a_1, y} p(b=a_y \mid a_0, a_1, y) = \frac{S + 4}{8}.
\end{equation}
Ref.~\cite{pawlowski2011semi} shows that secure key distribution is guaranteed when $P_B > P_E$, where $P_E$ is the eavesdropper's optimal guessing probability. Using results from guessing probability games, it is proved that security is ensured when $P_B > 0.8415$. 

%\subsection*{Wave-Particle Duality Quantities}

Our analysis also involves two operational quantities—input distinguishability $\mathcal{D}$ and interferometric visibility $\mathcal{V}$, which respectively characterize particle-like and wave-like behavior in binary interferometric setups. A central result of this work is the connection we establish between these quantities and the SDI witness, which will be discussed in the following sections.
Building on the framework introduced in Ref.~\cite{coles2014equivalence}, the input distinguishability $\mathcal{D}$ is defined in terms of the optimal probability of correctly guessing the which-path observable $Z$
\begin{equation}
\mathcal{D} := 2p_{\text{guess}}(Z) - 1.
\end{equation}
Operationally, this is estimated by blocking each path (upper $U$ or lower $L$) and measuring the difference in output detection probabilities $p_0$ and $p_1$:
\begin{equation}
\mathcal{D}_{U(L)} := \left[\frac{\abs{p_0 - p_1}}{p_0 + p_1}\right]_{\text{path \textit{U(L)} blocked}}, \quad
\mathcal{D} := \frac{1}{2}(\mathcal{D}_U + \mathcal{D}_L).
\end{equation}
The interferometric visibility $\mathcal{V}$ measures coherence via the guessing probability of the output observable $W$, optimized over all observables in the $XY$ plane~\cite{coles2014equivalence}
\begin{equation}
\mathcal{V} := \max_{W \in XY}[2p_{\text{guess}}(W) - 1].
\end{equation}
Operationally, $\mathcal{V}$ is determined in binary interferometers by varying a relative phase and recording the maximum and minimum detection probabilities at each output port
\begin{equation}
\mathcal{V}_j := \frac{p_j^{\text{max}} - p_j^{\text{min}}}{p_j^{\text{max}} + p_j^{\text{min}}}, \quad
\mathcal{V} := \frac{1}{2}(\mathcal{V}_0 + \mathcal{V}_1).
\end{equation}

\subsection{Mapping SDI witness via wave-particle duality scenarios}

\begin{figure*}[!htb]
\centering
\includegraphics[trim = {0.5cm 22.5cm 0.7cm 0.5cm},clip,width=\linewidth]{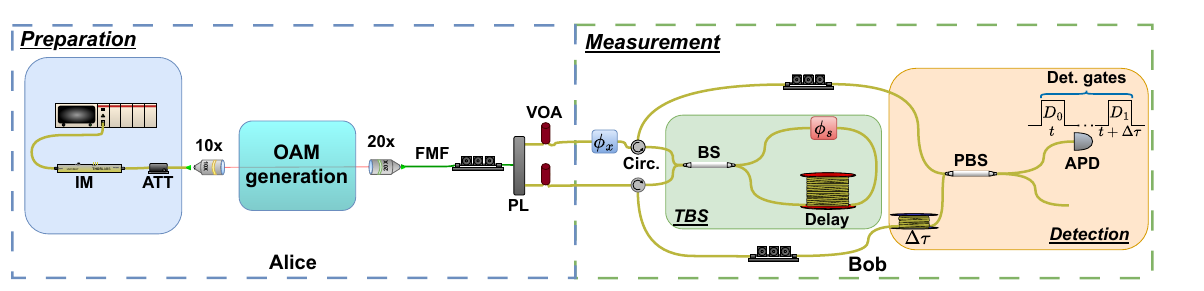}
\caption{\textbf{SDI witness implementation.} Alice prepares orbital angular momentum (OAM) weak coherent photonic states (WCS), produced from heavily attenuated optical pulses. The OAM states are coupled onto a few-mode fiber (FMF), which passes through a manual polarization controller wound with the FMF before being connected to a photonic lantern (PL), whose single-mode fiber outputs form a Mach-Zehnder interferometer. The PL splits the OAM state into its two linearly polarized components, $|LP_{11a}\rangle$ and $|LP_{11b}\rangle$, mapped to the two paths of the interferometer. Bob connects the two paths to a tunable beam splitter (TBS), comprised of a fiber-optical Sagnac interferometer, containing a phase modulator $\phi_s$ and a 300 m fiber-optical delay. When the two Sagnac counter-propagating paths recombine at the fiber beamsplitter (BS), the two outputs are split from the inputs using two optical circulators (Circ.). The two circulator outputs are then multiplexed in time with a fiber delay $\Delta\tau$ in the lower arm before being recombined at a polarization beam splitter (PBS), with the transmission of the two input paths maximized with manual polarization controllers. This allows the use of one single-photon detector, as the two-path outputs of the TBS become distinguishable due to the time separation $\Delta\tau$, effectively and thus matched to two detection times labeled as $D_0$ and $D_1$. An additional phase modulator $\phi_x$ is placed inside the interferometer to perform the visibility measurements. For additional details on the modeling and implementation, please refer to the main text as well as Methods Sections~\ref{m: modeling} and~\ref{m:exp}.   
}
\label{Figsetup}
\end{figure*}

We reinterpret this PM protocol as a wave-particle duality game, inspired by the MZI-based QKD protocol described in~\cite{Ardehali1996}. In this scenario, Alice's preparation involves deciding whether or not to include the first beam splitter $\text{BS}_1$, as depicted in Fig.~\ref{FigSDI}. 
It is worth mentioning that, in our experimental setup, this is represented by a photonic lantern (PL), please see Fig.~\ref{Figsetup} and Methods~\ref{m:exp} for further details. 
Specifically, particle-like behavior can only be observed in the MZI when $\text{BS}_1$ is removed (or when the paths are blocked after the first beam splitter); conversely, wave-like behavior is observed when $\text{BS}_1$ is included. Moreover, incorporating a final TBS in the setup, as shown in Figs.~\ref{FigSDI} and~\ref{Figsetup} and detailed in Methods~\ref{m: modeling}, allows one to interpolate between fully particle-like and fully wave-like behavior.
Analogously, without loss of generality, we associate even-parity inputs ($a_0 = a_1$) with particle-like experiments, and odd-parity inputs ($a_0 \neq a_1$) with wave-like experiments. Naturally, the output $b$ corresponds to the detector clicks ($D_0$ or $D_1$). Note that while $y$ determines the measurement setting, we consider the relative phase of the MZI, defined as a parameter $\phi_x$ , shown in Fig.~\ref{FigSDI}, to be later optimized. 

To address this parity-based structure within the binary MZI, we adopt the four-state preparation scheme used in both BB84 and~\cite{Ardehali1996}, given by
\begin{equation}
    \begin{gathered}
        \rho_{00} = \ket{0}\bra{0}, \quad
        \rho_{01} = \ket{-}\bra{-}, \\
        \rho_{10} = \ket{+}\bra{+}, \quad
        \rho_{11} = \ket{1}\bra{1}, \label{eq:states}
    \end{gathered}
\end{equation}
where $\ket{0}$ and $\ket{1}$ correspond to the two interferometric paths, while $\ket{\pm} = \frac{1}{\sqrt{2}}(\ket{0} \pm \ket{1})$ represent path superpositions for the cases where Alice included $\text{BS}_1$. Notice that the states in Eq.~\eqref{eq:states} follow uniquely from the MZI structure, and this particular choice of encoding ensures that the parity captures either particle-like or wave-like behavior. It is important to stress that this encoding is not restrictive, as any alternate bit-to-state assignment can be mapped to this one via relabeling. This structure provides a natural framework to study the trade-off between $\mathcal{D}$ and $\mathcal{V}$ within the SDI witness \eqref{eq:sdi_witness}.

In a PM scenario, the behavior is fully characterized by the prepared state $\rho_{a_0,a_1}$ and the measurement $M_y$ performed. Building on this and following the previously introduced operational definitions, we redefine $\mathcal{D}$ and $\mathcal{V}$ for each configuration $(a_0,a_1,y,b)$ as
\begin{align}
    &\mathcal{D}_{a_0,a_1, y}^{b} := 2p_{\text{path}, \phi_x}^{a_0, a_1, y} - 1, \label{eq:distinguishability} \\
    &\mathcal{V}_{a_0,a_1, y}^{b} := 2p^{b, a_0, a_1, y}_{\text{max}} - 1, \label{eq:visibility}
\end{align}
where
\begin{align}
    &p_{\text{path},\phi_x}^{a_0, a_1, y} \equiv P_{\phi_x}(b=a_0 | a_0, a_1, y), \label{eq:gpath} \\
    &p^{b, a_0, a_1, y}_{\text{max}} \equiv \max_{\phi_x} P_{\phi_x}(b | a_0, a_1, y). \label{eq:maxphi}
\end{align}
The form of Eq.~\eqref{eq:gpath} reflects that the path information is encoded in $a_0$, which is ensured by the encoding \eqref{eq:states}.

We are now equipped to connect the SDI witness $S$ to the wave-particle quantities. Due to the parity-based encoding, we define the path guessing probability as
\begin{equation}
    p_{\text{path}, \phi_x} = \frac{1}{4} \sum_{a_0, a_1, y} P_{\phi_x}(b = a_0 | a_0, a_1, y) \delta_{a_0 \oplus a_1 = 0}, \label{eq:totalpath}
\end{equation}
for the cases of particle-like experiments. 

Comparing \eqref{eq:suc_rac}, \eqref{eq:distinguishability} and \eqref{eq:totalpath}, the SDI witness $S_{\phi_x}$ can be rewritten as
\begin{align}
    S_{\phi_x} &= \sum_{a_0, a_1, y} \left( \frac{\mathcal{D}_{a_0, a_1, y} + 1}{2} \right) \delta_{a_0 \oplus a_1 = 0} \nonumber \\
    &\quad - 4 + p(0|01,0) + p(1|01,1) + p(1|10,0) + p(0|10,1).
\end{align}
Maximizing over the phase shift $\phi_x$ (which tunes the interference visibility), and using the definition of visibility in Eq.~\eqref{eq:visibility}, we finally obtain the core result
\begin{multline}
    \label{eq:maxphiS2}
    \max_{\phi_x} S_{\phi_x} = \frac{1}{2} \sum_{a_0, a_1, y} \mathcal{D}_{a_0, a_1, y} \, \delta_{a_0 \oplus a_1 = 0} 
    \\+ \frac{1}{2} \sum_{a_0, a_1, y, b} \mathcal{V}_{a_0, a_1, y}^b \, \delta_{a_0 \oplus a_1 = 1} \delta_{b = a_y}.
\end{multline}
It is important to emphasize that our analysis explores the independence of each experiment under the optimization of $\phi_x$, and that $\mathcal{D}_{a_0, a_1, y}$ remains phase-independent due to the parity encoding in~\eqref{eq:states}.

This decomposition reveals that the SDI witness $S$ directly quantifies the sum of optimal distinguishabilities and visibilities over the relevant configurations of the game. Importantly, this formulation remains valid under minimal assumptions about the underlying physical implementation, requiring only a single parameter to be maximized in the measurement setting.

Moreover, in Methods~\ref{M:Improving}, we present a detailed derivation of an upper bound on the eavesdropper’s success probability $P_E$ in terms of Bob’s average success probability $P_B$, refining the SDI security condition originally proposed in~\cite{pawlowski2011semi}. We obtain,
\begin{equation}
P_E \leq \frac{3 + \sqrt{1 - 2(2P_B - 1)^2}}{4}.
\label{eq:better_bound}
\end{equation}
This inequality implies that secure key distribution (i.e., $P_B > P_E$) is ensured whenever $P_B > 0.833$, thereby improving upon the previous threshold of $0.8415$. The full derivation is provided in Methods~\ref{M:Improving}. We now define the experimental context in which the SDI witness is assessed.

\subsection{Wave-particle duality experiments with a TBS}

We focus on the SDI witness within a particular WPD guessing game. This is achieved by analyzing a photonic interferometer equipped with an input $50:50$ beamsplitter (BS), a relative phase shift, and a TBS. We show how data obtained from two specific interferometer configurations—characterized by phase shifts $\phi_x$ and $\phi_x + \pi$—can be used to interpolate the SDI witness and demonstrate non-classicality and security conditions based on the interferometric quantities.

We now follow the setup depicted in Fig.~\ref{Figsetup}, and define Bob's measurements in the SDI guessing game as given by
\begin{align}
M_0(\phi_s,\phi_x) &= \cos{\phi_s}\sigma_z + \sin{\phi_s}\left(\cos{\phi_x}\sigma_x + \sin{\phi_x}\sigma_y\right), \nonumber \\
M_1(\phi_s,\phi_x) &=M_0(\phi_s,\phi_x+\pi), \label{eq:measurements}
\end{align}
where \(0 \leq \phi_s \leq \pi/2\), and \(\phi_x \in [0, 2\pi N]\) with \(N \in \mathbb{N}\), allowing phase scans over multiple cycles.
 These observables interpolate between wave-like and particle-like behavior depending on the relative phases. 
Under the symmetry of the interferometer and measurements, we obtain the relations
\begin{equation}
\begin{gathered}
E_{00,0} = E_{00,1}, \quad E_{11,0} = E_{11,1}, \\
E_{01,0} = E_{10,1}, \quad E_{10,0} = E_{01,1}.
\end{gathered}
\end{equation}
In addition to these symmetry relations, the constraints imposed by parity-obliviousness and blocking lead to the following normalization conditions
\begin{equation}
\begin{gathered}
E_{00,0} + E_{11,0} = 1, \\
E_{01,0} + E_{10,0} = 1.
\end{gathered}
\end{equation}
Here, parity-obliviousness ensures that the encoding does not reveal information about the parity bit $a_0 \oplus a_1$. Blocking is implemented by selectively disabling one of the interferometric paths, which allows marginal probabilities to be determined independently of interference, thereby isolating particle-like behavior.

Particularizing on the states and measurements discussed means exploiting the symmetries $\mathcal{V}_{0,1,y}^b = \mathcal{V}_{1,0,y}^b = \mathcal{V}$ and $\mathcal{D}_{0,0,y}^b = \mathcal{D}_{1,1,y}^b = \mathcal{D}$, which leads to the final compact expression
\begin{equation}
\max_{\phi_x} S_{\phi_x} = 2(\mathcal{D} + \mathcal{V}) .\label{eq:maxphiS_exp}
\end{equation}

In what follows, we present our experimental implementation of the SDI witness via complementary interferometric quantities.

\subsection{Experimental assessment of SDI witness via wave-particle quantities}
\begin{figure}
\centering
\includegraphics[width=1\columnwidth]{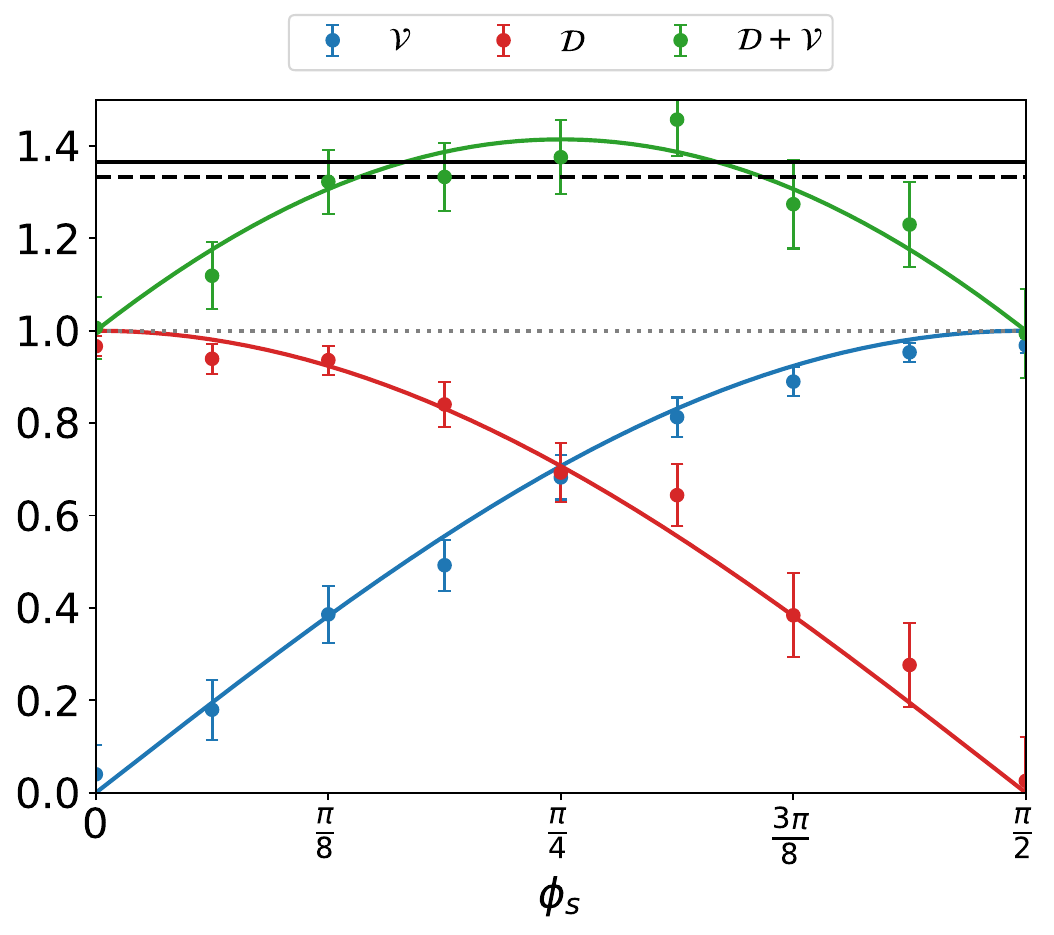}
\caption{\textbf{SDI witness via wave-particle relation}: The blue and red curves represent the average interferometric visibility $\mathcal{V}$ and average distinguishability $\mathcal{D}$, respectively. The green curve shows the SDI witness expressed as $S/2 = \mathcal{D} + \mathcal{V}$. The dotted black line marks the classical bound $\mathcal{D} + \mathcal{V} > 1$, while the dashed and solid lines indicate the two security thresholds: the previously known bound at $1.366$ and the improved bound at $1.332$. Each data point (circle) represents an experimental value, with error bars obtained via error propagation assuming Poissonian statistics for the detected event counts.}
\label{Fig2}
\end{figure}

 Our experimental setup, shown in Fig.~\ref{Figsetup} and detailed in Methods~\ref{m: modeling} and~\ref{m:exp}, is based on a modified fiber-optical Mach-Zehnder interferometer where the input beamsplitter is composed of a photonic lantern \cite{Birks_2015} which decomposes two-dimensional OAM states of light into its linearly polarized components in the following way: $|OAM_{+1}\rangle=1/\sqrt{2}(|LP_{11a}\rangle + i|LP_{11b}\rangle)$ \cite{Alarcon2021}. Within the two parallel paths of the interferometer, electro-optical attenuators and a phase modulator are placed, such that we can transform the original balanced superposition into the general path state $\alpha|LP_{11a}\rangle + e^{i\phi_x}\beta|LP_{11b}\rangle$, where the amplitude coefficients $\alpha$ and $\beta$ are adjusted with the attenuators and $\phi_x$ with the phase modulator. The interferometer's paths are then superposed on a $50:50$ fiber-optical beamsplitter, whose outputs form a fiber Sagnac interferometer, which works as a tunable beam splitter (TBS), dynamically adjustable by the internal phase modulator $\phi_s$. Two circulators are employed to route the outputs from the tunable beam splitters to a time-multiplexing fiber arrangement to map the two outputs to the two time slots $D_{0}$ and $D_{1}$, identifiable by one single-photon detector. Quantum states of light are prepared from a heavily attenuated laser beam and encoded onto the $|OAM_{+1}\rangle$ state with a spatial light modulator, which are coupled to a few-mode fiber, which is then connected to a photonic lantern decomposing the state onto the $|LP_{11a}\rangle$ and $|LP_{11b}\rangle$ modes, mapped respectively to the lower and upper paths of the interferometer. For more details on the experimental setup, we refer the reader to Methods~\ref{m:exp}.

We then proceed to measure the visibility of the $|OAM_{+1}\rangle$ state passing through the interferometer, by scanning $\phi_x$ for different settings of $\phi_s$ of the tunable beamsplitter corresponding to changing the measurement operator from particle to wave aspects, quantifying the SDI witness $S$. The distinguishability is measured in a similar fashion, with the difference that one of the arms of the interferometer is blocked with the electro-optical attenuators. Fig.~\ref{Fig2} shows that the experimental results match the theoretical framework connecting $S$ and the interferometric quantities $\mathcal{D}$ and $\mathcal{V}$, as developed in this work. Our results show that neither purely particle-like ($\mathcal{D} = 1$) nor purely wave-like behavior ($\mathcal{V} = 1$) is sufficient to violate the classical bound. This highlights the need for a balanced contribution from both aspects to reveal the non-classicality. As depicted in Fig.~\ref{Fig2}, the optimal violation occurs when $\mathcal{D} = \mathcal{V} = \sqrt{2}/2$, corresponding to a phase setting of $\phi_s = \pi/4$.

In Fig.~\ref{Fig2}, we compare two security criteria: the original bound of~\cite{pawlowski2011semi}, which, in terms of interferometric quantities, requires $\mathcal{D} + \mathcal{V} > 1.366$ for secure communication, and the improved bound derived in Methods~\ref{M:Improving}, which establishes the condition $\mathcal{D} + \mathcal{V} > 1.332$. This tighter threshold expands the range of parameters for which security can be certified in the experiment.

%{\it Discussion:}
\section*{Discussion}
\label{sec: di}

Recent developments have revealed that wave-particle duality can be understood as an optimized entropic uncertainty relation involving min- and max-entropies~\cite{coles2014equivalence,spegel2024experimental}, providing it with a clear operational meaning rooted in information theory. This perspective bridges a foundational concept—central to debates on the nature of light and the interpretation of quantum mechanics, including the famous EPR–Bohr debate—with the language of communication theory. It reinforces the view that quantum mechanics is fundamentally an informational theory. Building on this insight, our work establishes a direct theoretical connection between wave-particle duality—quantified by generic input distinguishabilities ($\mathcal{D}$) and interferometric visibilities ($\mathcal{V}$)—and a semi-device-independent (SDI) witness, linking duality to classical dimension bounds and security conditions relevant to quantum communication.

To illustrate our main result, we focused on a symmetric scenario involving two measurements realized through a tunable beam splitter (TBS) in the measurement setting. This choice allowed us to express the SDI witness in a simple form: $S = 2(\mathcal{D} + \mathcal{V})$. This relation offers a clear and physically meaningful way to interpret quantum behavior in communication tasks such as the quantum random access code (QRAC). Notably, it reveals that neither extreme particle-like behavior ($\mathcal{D} = 1$, $\mathcal{V} = 0$) nor extreme wave-like behavior ($\mathcal{D} = 0$, $\mathcal{V} = 1$) suffices to certify non-classicality in the SDI framework—highlighting that quantum advantage arises from a balanced interplay between these complementary aspects. 
This insight is consistent with previous findings, such as the demonstrated insecurity of the BB84 protocol in SDI scenarios~\cite{pawlowski2011semi}. Consequently, schemes that rely solely on extreme wave-particle behaviors—such as the delayed-choice cryptographic protocol~\cite{Ardehali1996}—also fail to provide SDI-certified quantum advantage.

We experimentally demonstrated the SDI certification of non-classicality in a prepare-and-measure scenario based on complementary measures of distinguishability $\mathcal{D}$ and visibility $\mathcal{V}$ of a quantum state. Our experiment was based on a dynamic fiber-optical platform in which we are able to continuously change between $\mathcal{D}$ and $\mathcal{V}$ measurement operators, and furthermore further demonstrates the experimental possibilities of performing quantum information processing using OAM states within an optical fiber platform. By expressing the SDI witness in terms of $\mathcal{D}$ and $\mathcal{V}$, our work connects foundational principles of quantum mechanics with operationally relevant quantities. It shows that wave-particle duality, when treated quantitatively and beyond its idealized extremes, can serve as a useful tool for certifying quantum behavior in practical communication tasks. This perspective not only deepens the conceptual understanding of duality-based experiments but also informs possible designs of SDI-secure quantum technologies.  An interesting direction for future work is to explore whether this interferometric decomposition offers advantages over standard SDI approaches under realistic conditions, particularly in the presence of no-click events and potential eavesdropping strategies.

\appendix

\section*{Methods}

\subsection{Improving the security bound in the SDI scenario}
\label{M:Improving}

The security framework of Ref.~\cite{pawlowski2011semi} establishes that in a 2-to-1 QRAC scenario, security is guaranteed if $ P_B > 0.8415 $, where $ P_B $ denotes the probability that Bob correctly infers the bit sent by Alice. In this section, we derive a tighter security bound on $ P_B $ refining the condition for secure SDI-QKD. 
Consider the success probability of Bob and Eve (the eavesdropper), trying to guess $a_0 \oplus a_1$. We have,
\begin{align}
    P_{B,E}(a_0 \oplus a_1) &\geq P_{B,E}(a_{0},a_{1})\\
    &\geq P_{B,E}(a_0)+P_{B,E}(a_1)-1.
    \label{eq:security_1}
\end{align}
where $P_{B,E}(a_0,a_1)$ (equivalently,$P_{B,E}(a_0 \cap a_1))$ is the likelihood that Bob and Eve guess both $a_0$ and $a_1$ correctly.
The second inequality follows from the identity
$P_{B,E}(a_0 \oplus a_1)=P_{B,E}(a_0 \cup a_1)-P_{B,E}(a_0\cap a_1)$ along with the upper bound $P_{B,E}(a_0 \cup a_1)=P_{B,E}(a_0)+P_{B,E}(a_1)-P_{B,E}(a_0\cap a_1)\leq 1$. 

For simplicity, we assume that $a_0$ and $a_1$ are uniformly distributed random variables, implying $P_{B,E}(a_0)=P_{B,E}(a_1)$, therefore \eqref{eq:security_1} becomes,
\begin{align}
   P_{B,E}(a_0 \oplus a_1) &\geq 2P_{B,E}(a_0)-1\\
    &= 2P_{B,E}-1.
    \label{eq:security_2}
\end{align}
where $P_{B,E}$ denotes Bob's average success probability. We consider the worst-case scenario where Bob and Eve collaborate under the constraints: $P_{B,E}(a_i) \geq  P_B(a_i)$ and $P_{B,E}(a_i) \geq  P_E(a_i)$. Using these constraints
\eqref{eq:security_2} becomes, 
\begin{equation}
P_{B,E}(a_0 \oplus a_1)\geq 2P_{E}-1
\label{eq:security_3}
\end{equation}
 To derive a better bound on $P_B$, we employ an inequality on the expectation values of  $a_0$, $a_1$, and $a_0 \oplus a_1$
\begin{equation}
 [E(a_0)]^2+[E(a_{1})]^2+[(E(a_0 \oplus a_1)]^2 \leq 1.\label{eq:bound_from_hyperbits}
\end{equation} 
Without loss of generality, one can write
$
    \rho = \frac{1}{2}(I + \textbf{n.}\boldsymbol{\sigma}),
$
where \textbf{n} is the Bloch vector and $\boldsymbol{\sigma} = (\sigma_x,\sigma_y,\sigma_z)$ are the Pauli matrices. Similarly, the measurement operators along the direction $\textbf{m}_i$ (where $i = \{0,1,2\}$) can be described by
$
    M_i = \frac{1}{2}(I + {\textbf{m}_i\boldsymbol{.\sigma}})
$.
Thus, the expectation value corresponding to each measurement $i$ is given by
\begin{equation}
    E_{i}  = 2 \textnormal{Tr}[\rho M_i]-1 = (\textbf{n.m}_i).
\end{equation}
Now, we consider the expectation values $E_i(f_j)$ of the function $f_j$, estimated from Bob's collected statistics, where $f_j := \{a_0, a_1, a_0 \oplus a_1 \}$. In this case, we have:
\begin{equation}
\begin{gathered}
    \sum_{i=a_0, a_1, a_0 \oplus a_1}[E_i(f_i)]^2 = [E(a_0)]^2+[E(a_1)]^2+[E(a_0 \oplus a_1)]^2\\
    =\sum_i (\textbf{n.m}_i)^2 \leq |n|^2\sum_i\textbf{m}_i.
\end{gathered}
\end{equation}
Since $\sum_i\textbf{m}_i \leq1$, and $|n|^2 \leq1$, we have \eqref{eq:bound_from_hyperbits} as a result.

Since $E(a_0 \oplus a_1)=2P_{B,E}(a_0\oplus a_1)-1$, $E(a_0)=2P_{B,E}(a_0)-1$ and $E(a_1)=2P_{B,E}(a_1)-1$, \eqref{eq:bound_from_hyperbits} simplifies to,
\begin{align}
P_{B,E}(a_0 \oplus a_{1}) &\leq \frac{1+\sqrt{(1-2(2P_{B,E}-1)^2)}}{2}\\
&\leq \frac{1+\sqrt{(1-2(2P_{B}-1)^2)}}{2}
\end{align}

Using \eqref{eq:security_3}, we derive an upper bound on $P_E$
\begin{equation}
   P_E \leq \frac{3+\sqrt{(1-2(2P_{B}-1)^2)}}{4}. \label{eq:tighter_bound}
\end{equation}
This implies that $P_B>P_E$ as long as $P_B>0.833$. In the following, we present our experimental verification of the SDI witness in terms of interferometric distinguishability and visibility.

\subsection{Interferometric model}
\label{m: modeling}

The interferometric setup under analysis is illustrated in Fig.~\ref{Figsetup}. Each optical component is modeled by its corresponding matrix operator
\begin{equation}
BS_1 = \frac{1}{\sqrt{2}} \begin{pmatrix}
1 & i \\
i & 1
\end{pmatrix}, \quad
BS_2 = \frac{1}{\sqrt{2}} \begin{pmatrix}
i & -1 \\
-1 & i
\end{pmatrix},
\end{equation}
\begin{equation}
PM_1 = \begin{pmatrix}
1 & 0 \\
0 & e^{i\phi_x}
\end{pmatrix}, \quad
PM_2 = \begin{pmatrix}
1 & 0 \\
0 & e^{i\phi_s}
\end{pmatrix}.
\end{equation}
By applying these transformations, the output state becomes
\begin{equation}
\begin{aligned}
\ket{\psi} = \frac{1}{2\sqrt{2}} &\left[i\left((1+e^{i \phi_s})+e^{i \phi_x}(1-e^{i \phi_s})\right)\ket{D_0} \right.\\
&\left. + \left((e^{i \phi_s}-1)-e^{i \phi_x}(1+e^{i \phi_s})\right)\ket{D_1}\right]
\end{aligned}
\end{equation}
The detection probabilities at detectors $D_0$ and $D_1$ are then given by
\begin{align}
p_0 &= |\bra{D_0}\psi\rangle|^2 = \frac{1}{2}(1+\sin{\phi_x}\sin{\phi_s}) \label{eq:p1}\\
p_1 &= |\bra{D_1}\psi\rangle|^2 = \frac{1}{2}(1-\sin{\phi_x}\sin{\phi_s}) \label{eq:p2}
\end{align}
Maximizing and minimizing these probabilities allows us to extract the interferometric visibility:
\begin{equation}
\mathcal{V} = \sin{\phi_s} \label{eq:V}
\end{equation}

To compute the distinguishability, we consider blocking one of the interferometric paths. If the upper path is blocked, the resulting state is
\begin{equation}
\ket{\psi} = \frac{1}{2}\left[i(1+e^{i\phi_s})\ket{D_0} + (1-e^{i\phi_s})\ket{D_1}\right]
\end{equation}
Since the phase $\phi_x$ becomes irrelevant when only one path is available, the distinguishability becomes
\begin{equation}
\mathcal{D} = \cos{\phi_s} \label{eq:D}
\end{equation}

\subsection{Experiment}
\label{m:exp}

The experimental setup resorts to a fiber-optical Mach-Zehnder interferometer where the input beamsplitter is composed of a photonic lantern, which takes a few-mode fiber (FMF) as an input, and splits the corresponding linearly polarized (LP) modes supported by the FMF. The FMF employed is a 3-mode fiber from OFS, supporting the fundamental Gaussian $|LP_{01}\rangle$ mode, as well as $|LP_{11a}\rangle$ and $|LP_{11b}\rangle$. In our case we only use the two higher-order ones, as they form the linear decomposition of the $|OAM_{+1}\rangle$ mode. The $|OAM_{+1}\rangle$ state itself is generated from heavily attenuated laser pulses generated from an electro-optical LiNbO$_3$ fiber pigtailed amplitude modulator connected to the output of a direct feedback (DFB) laser diode emitting at the wavelength of 1546 nm. The amplitude modulator is driven from a signal generator creating 40 ns wide optical pulses at 150 kHz repetition rate. The attenuator is adjusted such that there is an average photon number per pulse of $\mu =0.2$ following the poissonian distribution $P(n)=\mu^ne^{-\mu}/n!$ just before the tunable beam splitter, which is where the measurement operation takes place. 

The OAM state is generated with a free-space setup based on a spatial light modulator (SLM). A 10$\times$ objective is used to collimate the light beam before passing through a polarizer to prepare a horizontal polarization state before impinging onto the SLM, which has an appropriate forked hologram displayed, creating the $|OAM_{+1}\rangle$ on the first diffraction order. a 4f optical system is employed together with a pinhole spatially filtering the reflected beam from the SLM, and then coupled into the FMF with a 20$\times$ objective. A manual polarization controller with the FMF wound through the paddles is used to compensate for the disturbances of spatial distribution on the FMF, ensuring optimal splitting of the $|LP_{11a}\rangle$ and $|LP_{11b}\rangle$. The employed photonic lantern (Phoenix Photonics) gives around 16 dB of extinction ratio between the used outputs, and around 5 dB of average insertion loss. 

Following the interferometer, the tunable beamsplitter is implemented through a fiber-optical Sagnac interferometer. The two counter propagating directions in the Sagnac loop recombine back at the input beamsplitter, and depending on the relative phase difference $\phi_s$ between the two directions, the detection probability at detectors $D_0$ and $D_1$ (following the optical circulators) is proportional to $cos^2(\phi_s)$ and $sin^2(\phi_s)$ respectively. An optical fiber delay inside the Sagnac loop (300 m long for convenience) is employed to ensure that the modulator $\phi_s$ only acts in one of the opposing internal propagation directions. Following the circulators we employ one single-photon detector (SPD) module based on an avalanche InGaAs photodiode in gated mode (IdQuantique id210). In order to be able to detect both outputs with a single-detector, we resort to a time-multiplexing scheme, where a $\Delta\tau = 1250$ ns optical fiber delay is inserted onto one of the outputs, and then both are recombined with a polarizing beam splitter (PBS) and the output connected to the SPD. This time delay was chosen based on the internal deadtime of the detector, which is set to minimize false counts from afterpulsing events. Manual polarization controllers are employed to maximize transmission through both inputs of the PBS. This scheme effectively maps the two outputs from the interferometer onto a single detector, with the drawback of limiting the repetition rate of the experiment.

\section*{Acknowledgements}
This work was supported by Zenith Linköping University and the Wallenberg Center for Quantum Technologies. This work was partially supported by the Foundation for Polish Science (IRAP project, ICTQT, contract No. MAB/218/5, co-financed by EU within the Smart Growth Operational Programme). M.P., C.R., and P.R.D. acknowledge support from the NCN Poland, ChistEra-2023/05/Y/ST2/00005 under the project Modern Device Independent Cryptography (MoDIC). A.C. acknowledges financial support by NCN grant SONATINA 6 (contract No. UMO-2022/44/C/ST2/00081). This work is partially carried out under IRA Programme, project no. FENG.02.01-IP.05-0006/23, financed by the FENG program 2021-2027, Priority FENG.02, Measure FENG.02.01., with the support of the FNP.

%##########################################################################
%##########################################################################
%##########################################################################

\bibliography{bib}
%\bibliography{bibliography}

\end{document}